\documentclass[preprint,showpacs,preprintnumbers,amsmath,amssymb]{revtex4}

\usepackage{graphicx}
\usepackage{dcolumn}
\usepackage{bm}

\begin{document}

\title{Interlinguistic similarity and language death dynamics}

\author{Jorge Mira}
\email{fajmirap@usc.es}
\affiliation{
Departamento de F\'\i{}sica Aplicada, Universidade de Santiago de Compostela
E-15782 Santiago de Compostela, Spain}
\author{\'Angel Paredes}
\email{angel@fpaxp1.usc.es}
\affiliation{
Departamento de F\'\i{}sica de Part\'\i{}culas, Universidade de Santiago de Compostela
E-15782 Santiago de Compostela, Spain
}

\date{\today}

\begin{abstract}
We analyze the time evolution of a system of two coexisting languages (Castillian Spanish and Galician, both spoken in northwest Spain) in the framework of a model given by Abrams and Strogatz [Nature {\bf 424}, 900 (2003)]. It is shown that, contrary to the model's initial prediction, a stable bilingual situation is possible if the languages in competition are similar enough. Similarity is described with a simple parameter, whose value can be estimated from fits of the data.\end{abstract}

\maketitle

There is hard evidence that the number of languages in the world is shrinking. Of the roughly 6500 languages now spoken many of them are endangered or on the brink of extinction. The situation has attracted the interest of many researchers in the field of complex adaptive systems, who have analyzed language dynamics \cite{Gell-Mann,Abrams}. Among them, Abrams and Strogatz \cite{Abrams} have recently proposed a simple and creative model for the dynamics of language death. In it, they consider a system of two languages competing with each other for speakers, where the attractiveness of a language increases with both its number of speakers and its perceived status (a parameter that reflects the social and economic opportunities afforded to its speakers).

This model, which satisfactorily fits historical data on the decline of Welsh, Scottish Gaelic, Quechua and other endangered languages, predicts that one of the competing languages will inevitably die out. The bilingual societies that do in fact exist are thought by the authors to be, in most cases at least, unstable situations resulting from the recent merging of formerly separate communities with different languages. Here we suggest that stable bilingualism may be possible, and that whether it occurs or not may depend on the degree of similarity between the two competing languages. 

\begin{figure}[t]
\includegraphics[width=0.5\textwidth]{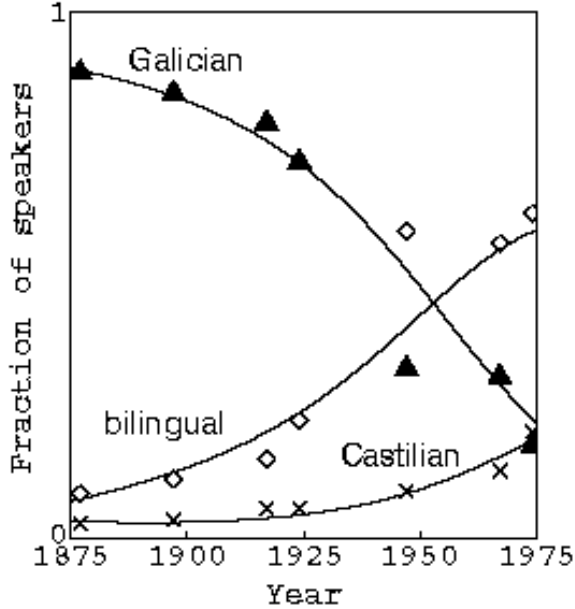}
\caption{Fraction of speakers vs. time in Galicia. Data from the Real Academia Galega (Royal Academy of Galicia) \cite{RAG}. We consider here only data obtained for dates prior to 1978, when the status of Galician began to increase \cite{Euromosaic} as the result of political support that included its becoming co-official with Castilian. The data for 1945 may nevertheless be considered as anomalous, since the use of Spanish languages other than Castilian was severely repressed following the end of the Spanish Civil War in 1939. The smooth curves are the result of fitting our modified Abrams-Strogatz model; the parameters of the fitted model are $a$ = 1.50, $s_{Galician}$ = 0.26, $c$ = 0.1 and $k$ = 0.80.}

\label{un}
\end{figure}

In Abrams and Strogatz' model, bilingualism is strictly societal: two monolingual groups coexist without there being any bilingual individuals. This model cannot account for situations such as that of Galicia (northwest Spain), where the outcome of competition between two Romance languages, Galician (the low-status language throughout the period considered) and Castilian Spanish (the high-status language), has been the existence of a bilingual majority alongside Galician and Castilian monolingual minorities \cite{RAG} (Fig. \ref{un}).

It would appear to be relevant to this question that in the cases considered by Abrams and Strogatz the admissibility of ignoring bilinguals is made plausible {\it by the great dissimilarity between the competing languages}: conversation is impossible between Quechua and Spanish, Welsh and English, and Scottish Gaelic and English monolingual speakers. By contrast, Galician and Castilian are very similar: both arose from Latin, and limited conversation is possible between monolingual speakers of Galician and monolingual speakers of Castilian. Moreover, the similarity of their grammar and vocabulary makes it easy to learn one of these languages when the other one is known. It therefore seems possible that the emergence and survival of a socially significant bilingual group may depend on the similarity between the competing languages. 

The Abrams-Strogatz model can be generalized to incorporate this notion as follows. Denoting by $X$, $Y$ and $B$ the subsets of the population that are monolingual in language X, monolingual in language Y, and bilingual, respectively, and by $x$, $y$ and $b$ the fractions of the population that belong to these groups ($x+y+b = 1$), the rate of change of $x$ is given by 

\begin{equation}
\frac{dx}{dt} = yP_{YX} + bP_{BX} - x(P_{XY} + P_{XB})
\end{equation}
(with analogous equations for $dy/dt$ and $db/dt$), where $P_{\alpha \beta}$ is the fraction of group $\alpha$ that transfers to group $\beta$ per unit time. Like Abrams and Strogatz, we take the probability per unit time of a member of $X$ beginning to speak Y to be given by a function of the form $cs_Y(1-x)^a$, where $s_Y$ is the relative status of Y ($0 \leq s_Y \leq 1$; $s_X = 1 - s_Y$) and $c$ and $a$ are constants; but we split this probability between the probability of becoming bilingual and the probability of beginning to use only Y: 

\begin{equation}
P_{XB} = cks_Y(1-x)^a
\end{equation}
and

\begin{equation}
P_{XY} =c(1-k)s_Y(1-x)^a
\end{equation}
where the parameter $k$ ($0 \leq k \leq 1$) reflects the ease of bilingualism and hence, according to our hypothesis, the similarity of the two languages. $k=0$ would represent situations where conversation is impossible between monolingual speakers (like the cases chosen by Abrams and Strogatz) and $k=1$ implies X=Y. It is worth mentioning that the concept of distance or similarity between languages has already been described from the theoretical point of view, for example by Nowak {\it et al.} \cite{Nowak}; nevertheless, its calculation in practical cases is extremely difficult. Similarly, 

\begin{equation}
P_{YB} = cks_X(1-y)^a
\end{equation}
and

\begin{equation}
P_{YX} =c(1-k)s_X(1-y)^a
.
\end{equation}

For transfers from $B$ to $X$ we take $P_{BX}  = P_{YX}$ (since both $B$-to-$X$ and $Y$-to-$X$ transfers involve loss of language Y, which mainly happens after death of the speaker), and similarly $P_{BY}  = P_{XY}$.  We thus obtain a pair of coupled differential equations for $x$ and $y$:

\begin{eqnarray}
\frac{dx}{dt} &=& c [(1-x)(1-k)s_X(1-y)^a\nonumber\\  && -  x(1-s_X)(1-x)^a]
\\
\frac{dy}{dt} &=& c [(1-y)(1-k)(1-s_X)(1-x)^a\nonumber\\  && -  y s_X(1-y)^a]
\label{eq:one}
\end{eqnarray}
which, obviously, reduces to the Abrams-Strogatz equation when $k = b= 0$. 

In Fig. \ref{un} it is shown that the modified model fits successfully the data and yields, as expected, a high similarity among both languages. An important fact is that we have found, from empirical observations of our numerical studies, that for every value of $s_X$ there exists $k_{min}(s_X,a)$, such that the language with less status dies out for all $k <  k_{min}$, but for all $k > k_{min}$ both groups $B$ and $X$ survive. This can be understood because for $k <  k_{min}(s_X,a)$ there exist values $0 < x_f, y_f < 1$ for which the right hand sides of Eqs. 6 and 7 vanish and towards which there is also an asymptotic tendency. For $k <  k_{min}(s_X,a)$, the only solutions of $\frac{dx}{dt} = \frac{dy}{dt}= 0$ are given by $x=1, y=0$ (unstable point provided $s_X <1/2$) and $x=0, y=1$.

In summary, we have shown that, in a model of competing languages, bilingualism is possible. Similarity appears to be the key factor that enables the stability of bilingualism. Our modification of the Abrams-Strogatz model allows also to estimate a coefficient of similarity between two languages.

\begin{acknowledgments}
We wish to acknowledge the help of Professors Ant\'on Santamarina Fern\'andez and Manuel Gonz\'alez Gonz\'alez, from the Departamento de Filolox\'\i{}a Rom\'anica and Instituto da Lingua Galega, Universidade de Santiago de Compostela, members of the Real Academia Galega (Royal Academy of Galicia).
\end{acknowledgments}

\end{document}